\documentclass{article}
\usepackage{amssymb}
\usepackage[normalem]{ulem}  
\usepackage{color}
\usepackage{graphicx}
\usepackage{amssymb}
\usepackage{epstopdf}
\def\Frac#1#2{\frac{\displaystyle #1}{\displaystyle #2}}

\begin{document}
\title{On the boundary conditions for the neutron transport equation}
\author{P. Saracco$^{1,2}$, N. Chentre$^1$, S. Dulla$^{3,4}$, P. Ravetto$^{3,4}$\\
\small
$^1$ I.N.F.N. - Sezione di Genova (Italy), Via Dodecaneso, 33 - 16146 Genova (Italy)\\ 
\small
$^2$ Centro Fermi, Piazza del Viminale, 1 - 00184 Roma (Italy)\\
\small
$^3$ Politecnico di Torino, Dipartimento Energia, \\ \small Corso Duca degli Abruzzi, 24 - 10129 Torino (Italy)\\
\small 
$^4$ I.N.F.N. - Sezione di Torino - Via P. Giuria, 1 - 10125 Torino (Italy)
\normalsize
}
\date{}
\maketitle
\begin{abstract}
The solution of the linear transport equation used for the study of neutral particle fields requires the imposition of appropriate boundary conditions. The choice of the conditions to impose for an infinite medium is not straightforward. The question has been given different formulations in the literature with various justifications based on some physical reasoning. Some aspects of the question are here analysed, from both the mathematical and the physical point of view. It is concluded that the inspiring golden rule should be the establishment of conditions that do not require any reference to the properties of the specific medium being considered for their justification.
\end{abstract}
\vskip0.15truecm
In this Letter we try to elucidate some points concerning the appropriate boundary conditions to be imposed on the solution of the Boltzmann equation for neutrons, particularly in the case of an infinite system. Although this equation has now been deeply studied by mathematical-physicists \cite{casezweifel} and widely used for applications by nuclear engineers \cite{henry}, the issue of boundary conditions seems not to have yet found a generally accepted formulation, in particular for problems in infinite media. 

The interest in the investigation of problems in the infinite domain is due to the fact that physicists often pretend that infinite systems show simplifications which are not present in the case of a finite system, although it is a pure abstraction \cite{Davison1957}. However, the extension of the spatial domain to infinity introduces some subtleties that deserve an in-depth analysis and determine the opportunity to clarify some important mathematical concepts. 

The linear Boltzmann equation concerns the mathematical description of transport of neutral particles: it is a classical particle balance in the phase space based on a statistical approach for the evaluation of the contribution of collisions yielding the behaviour of the statistical (mean value) of the number density of particles or, more commonly, of the corresponding particle angular flux \cite{Davison1957}; the linearity derives from assuming no direct interaction between particles in the interaction kernel. It constitutes, therefore, the fundamental model for the neutronics of nuclear reactors, although it may be applied to other physical problems, e.g. the physics of photon propagation.  It is quite natural to consider such equation for a finite system  - for simplicity assumed to be convex -  facing vacuum: then it is immediate to conclude that, since no particles can enter back into the system from vacuum, the angular flux $\psi$ must always vanish at all points $\vec r_S$ at the external bounding surface with outgoing normal $\hat n_S$, for all incoming directions, i.e.: 
\begin{equation}
\psi\left(\vec r_S, \hat\Omega, E,t\right)=0\qquad {\rm if} \qquad \hat\Omega\cdot\hat n_S<0\,. 
\label{eqnBoltzCond}
\end{equation}

Some difficulties are encountered when trying to push the bounding surface to infinity. This process seems to be altogether logically acceptable whenever the sources are localized. A discussion on this topic can be found in the classical book by Davison \cite{Davison1957}: his conclusion is that "{\em ... the condition at infinity should always imply that the number of neutrons coming directly from infinity is zero}", even if it is not clear how this physical statement should be translated into some consistent mathematical form. 

Mainly during the sixties, many authors tackled, often indirectly, this problem: we are not interested here in a complete analysis of the extended and rich bibliography, we simply summarize it remarking that a general consensus seems to arise on the fact that \cite{GibbsLarsen1977,DudMat79}:
\begin{equation}
\lim_{\left|\vec r\right|\to+\infty}\psi\left(\vec r,\hat\Omega,E,t\right)=0\,.
\label{eqn:genass}
\end{equation}
For instance, Case states that "{\em in general some conditions at infinity will be needed to make}" the solution "{\em unique. Here let us assume that vanishing at infinity is an appropriate requirement}" without any further explanation \cite{CASE}.
It is worth remarking that this conclusion is often hidden into some assumption on the space of functions in which the solution is searched; for instance, it is often assumed that in finite slab geometry the solution should belong to the Hilbert space of square integrable functions over $\left[-a,a\right]$ \cite{Mika1966}, which clearly is coherent with Eq. (\ref{eqn:genass}) in the limit $a\to\infty$.

On physical grounds we believe that the assumption (\ref{eqn:genass}) is too strong and not fully justified. In the absence of singular sources, the most general requirement is that the number density of particles never diverges, i.e. $\psi\left(\vec r, \hat\Omega, E,t\right)$ is bounded. This is quite obvious and generally accepted for finite systems. However, if we let the dimensions of the system tend to infinity, this seems to indicate that the functional space where to search for solutions is the space of bounded functions over $\mathbb R^3$, as far as the space dependence is concerned. The boundedness (and not the vanishing) of the solution is sometimes tacitly assumed when solving problems for some specific geometry, as is the case of the infinite slab problem for which it is obviously assumed the flux to be limited (and not vanishing) approaching infinity along the cross coordinates.

Is this last and weaker assumption coherent with the condition expressed by Eq. (\ref{eqnBoltzCond})? Various arguments can be used, but the simpler one is to consider the vacuum external to a finite system as a medium by itself, where clearly neutrons propagate along straight lines. Therefore,  neutrons coming from inside the system can surely propagate freely and no neutron can return back into the system coming directly from infinity. This is also coherent with the quoted requirement assessed by Davison. We remark that this is also obviously coherent with the form the transport equation takes in the vacuum
\begin{equation}
\left(\Frac{1}{v}\Frac{\partial}{\partial t} -\hat\Omega\cdot\vec\nabla\right)\psi\left(\vec x,\hat\Omega,E,t\right)=0,
\label{eqn:VacEqn}
\end{equation}
which naturally is obtained from the transport equation when setting all the macroscopic cross sections to zero. Naturally one requires also - as an interface condition - the continuity of the neutron angular flux on the convex surface delimiting the system from the surrounding vacuum, in the absence of sources on such a surface. 

Another instructive example can be made referring to the solution of the steady-state diffusion equation (which is well-known to be an approximation of the transport equation) in plane geometry, which takes the form of a decaying exponential in a purely diffusive medium, as $\exp(-\left|x\right|/L)$: obviously vacuum is simulated by a medium in which $L$ approaches $\infty$, and, hence the solution approaches a constant finite value, no matter how large the value of $x$ may be. In fact, Meghreblian and Holmes, in connection with diffusion theory, simply require the finiteness of the neutron flux \cite{megh}. 

Our interest is here restricted only to the form that the condition to be imposed to the solution of the neutron transport equation could assume at infinity, not in discussing the physical validity of the model itself: it should be questioned, in fact, if neglecting neutron-neutron interaction is physically reasonable in vacuum - where it is the only surviving interaction. Of course rigorously it is not, but this kind of problems is outside the scope of the present Letter: here we simply assume the validity of the model (the linear transport equation), independently from the arguments justifying it, and try to determine the most proper boundary conditions. Just to mention it, very far from the sources it is also questionable if fluctations around the mean number of neutrons in a small volume are negligible or not. But this kind of considerations opens a wider field of investigations.

It should also be noted that for plane one-dimensional problems the requirement of the finiteness of the particle angular flux at infinity in vacuum is effective, because if some particles are present and travelling in some outward direction at the physical surface of the material body, then they can be found after some time at whichever arbitrary distance in the same direction. For two- and three-dimensional problems, in the case of an isotropic source, the particle density tends naturally to zero for geometrical reasons, as one can easily see by using, for instance, the Peierls' integral form of the transport equation \cite{Davison1957}. However, this may not be true for a problem involving a collimated, or, more generally, non-isotropic source.

At last, it is worth to make a remark with regards to the conditions to be applied when considering the adjoint transport equation that is a powerful tool in many applications \cite{lewins}. Its solution can be physically interpreted as the particle importance. Based on physical reasoning, it is obvious that for its solution a vanishing condition cannot be applied at infinity in vacuum, since particles flying towards the medium being considered have a chance to contribute to collisions and, hence, retain a certain importance.

A fundamental requisite must be kept in mind when choosing the appropriate form of the boundary or limiting conditions to be imposed to the solution of a given equation: such conditions shall not depend on any specific values the coefficients or parameters appearing in the formulation of the equation might assume (such as the cross sections in this specific case): the vanishing of the angular flux at infinity is sometimes assumed as a consequence of the fact that in any "physical medium" some absorption is present, so forcing the solution to vanish at infinite distance from the sources. We believe that this is a not the appropriate way of reasoning, as one can always imagine for instance - even if as a {\it gedankenexperiment} - a purely scattering infinite medium. The mathematical conditions assumed for the solution at the boundary should be instead universally valid.

In conclusion the proper boundary conditions to be imposed on the solution of linear Boltzmann equation in the case of an infinite spatial domain is the boundedness of the angular flux, both in the direct and in the adjoint formulations, from which the appropriate functional space where to search for the solution follows in a straightforward manner. It is worth to remark that this conclusion, which naturally descendss from the  arguments presented, cannot be retrieved directly from eq. (\ref{eqnBoltzCond}) because for the case of an infinte spatial domain the concept of outgoing normal is not defined.

\end{document}